# Strong THz and Infrared Optical Forces on a Suspended Single-Layer Graphene Sheet

S. Hossein Mousavi,[1,*] Peter T. Rakich,[2] and Zheng Wang[1,†]

[1]Microelectronics Research Center, Cockrell School of Engineering, University of Texas at Austin, Austin, TX 78758 USA

[2]Department of Applied Physics, Yale University, New Haven, CT 06520 USA

ABSTRACT: Single-layer graphene exhibits exceptional mechanical properties attractive for optomechanics: it combines low mass density, large tensile modulus, and low bending stiffness. However, at visible wavelengths, graphene absorbs weakly and reflects even less, thereby inadequate to generate large optical forces needed in optomechanics. Here, we numerically show that a single-layer graphene sheet is sufficient to produce strong optical forces under terahertz or infrared illumination. For a system as simple as graphene suspended atop a uniform substrate, high reflectivity from the substrate is crucial in creating a standing-wave pattern, leading to a strong optical force on graphene. This force is readily tunable in amplitude and direction by adjusting the suspension height. In particular, repellent optical forces can levitate graphene to a series of stable equilibrium heights above the substrate. One of the key parameters to maximize the optical force is the excitation frequency: peak forces are found near the scattering frequency of free carriers in graphene. With a dynamically controllable Fermi level, graphene opens up new possibilities of tunable nanoscale optomechanical devices.

**Keywords:** Graphene, radiation pressure, Coulomb-Lorentz force, optomechanics, perturbation theory

Graphene interacts with light in unusual ways: while it is virtually invisible to light polarized normal to its surface, it interacts strongly with THz and mid-infrared light polarized parallel to its surface, even though it is only a monolayer of atoms[1–9]. This interaction is highly frequency-dependent due to the interband and intraband transitions of the free carriers in graphene[2,10–12]. The collective oscillations of these free carriers can also produce nanoscale plasmons, enabling the confinement of light to dimensions much smaller than the light wavelength[13–17]. Meanwhile, graphene possesses several attractive mechanical properties[18]: as a single atomic layer, graphene has very low mass density; while a free flake of graphene can be easily bent[19,20], it has an exceptionally stiff in-plane Young's modulus ~1 TPa[21]. On the surface with another material, it behaves like a fluidic interface and readily conforms, thanks to a remarkably strong adhesion with large Van der Waals forces on the order of 1 GPa[22,23].

Exploiting these unusual optical and mechanical properties together, one can envision intriguing uses of graphene in optomechanics. In recent years, many 3D micro- and nano-structures haven been explored in tailoring and enhancing optical forces in a wide range of applications, including reduced Brownian motion of atoms and resonators through optical cooling[24–28], tunable integrated optics[29–33], enhanced nonlinear parametric processes[34,35] and stimulated phonon generation[36]. Moving from 3D thin-film structures towards a 2D graphene sheet further reduces the mass density to the extreme, and can translate into larger optomechanical coupling and accelerations, resulting in better performance and lower power consumption. Using optical forces to mechanically actuate graphene is also practically attractive: comparing to photo-thermal and electrostatic transductions[18], direct actuation by optical forces provides a unique combination of capabilities including ultrafast modulation speed (GHz and beyond), high spatial resolution, and all-optical reconfigurability through spatially interfering multiple incident beams.

However, using graphene as an optomechanical material may not be entirely intuitive, and one needs to identify conditions that allow strong optical forces on graphene to emerge. Considering the widely-known low reflectivity (~0.01%) and low absorption (~2%) of graphene[37] at visible wavelengths, one would expect very weak optical forces from both absorption and scattering. This issue is resolved when we shift our attention to THz and infrared frequencies between 0.1 and 300 THz, a frequency range in which graphene reflects fairly strongly. Generally, one can optically excite graphene either through far-field illumination[38] or through near-field approaches[4,13,14,39–41]. In particular, near-field excitation allows one to access graphene plasmons with nanoscale confinement and strong field enhancement. However, graphene plasmons are also associated with practical challenges, including high propagation losses and demanding coupling mechanisms. In this paper, we instead focus on a simple system consisting of a single-layer graphene sheet suspended above a substrate and illuminated by far-field plane waves (Fig. 1). The optical forces can still be enhanced to large amplitudes, thanks to Fabry-Perot resonances formed between the graphene sheet and the substrate. And despite its seeming simplicity, such a system exhibits rich optomechanical behaviors ranging from widely tunable optical forces both in amplitude and direction (attractive/repellent) to all-optical levitation and precise positioning of the graphene layer. A key question we address here is how one can accomplish large optical forces in this simple system[42], given the control on the illumination condition and the dielectric environment.

The Article is structured as follows. We first use perturbation theory to decompose optical forces on graphene in near- and mid-infrared ranges by their physical origin: ohmic loss and kinetic inductance. We then expand our investigation to the entire THz and infrared range, in which graphene can significantly alter its surrounding fields. We study in detail the effect of excitation frequency, suspension height, and substrate permittivity. Four distinct types of substrates are considered: high-index dielectrics, metals, highly-doped semiconductors, and free space (free-standing graphene).

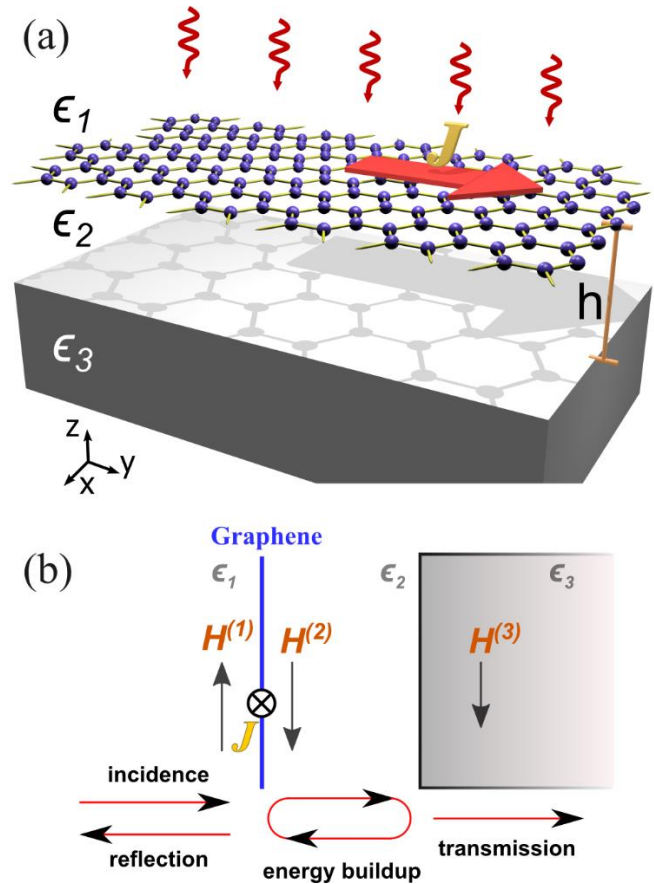

Figure 1. (a) Schematic of a one-dimensional optomechanical system with a graphene layer suspended in air ($\epsilon_1 = \epsilon_2 = 1$) at a distance $h$ from a substrate (partial mirror) with a relative permittivity of $\epsilon_3$. The direction of propagation is shown by the arrows. $J$ is the induced surface current. (b) Side view of the system, highlighting the field discontinuity and the circulating power between graphene and the substrate.

**Perturbative Treatment Using the Coulomb-Lorentz Force Law**

We begin with the near-infrared spectral range, in which graphene is a weak perturbation to its surrounding fields, and apply the law of the Coulomb-Lorentz force to reveal the major underlying sources of optical forces. Although optical forces can be exactly evaluated from the Maxwell stress tensor (see Supporting Info S1 for the closed form formulation), the knowledge of the main contributors to the optical forces is a significant insight that will be later extended to the more general non-perturbed regime.

In classical electromagnetics, graphene can be generally viewed as an infinitesimally-thin film -- though it is capable of producing significant perturbation -- with a complex-valued surface conductivity. Optical force on graphene is simply the electromagnetic force exerted on the surface currents and surface charges carried by this thin film. However, using the Coulomb-Lorentz force law to evaluate the force is only accurate when graphene weakly perturbs the system, because a graphene sheet in any optical system introduces discontinuities to the electromagnetic fields surrounding it, and renders the field terms in the Coulomb-Lorentz force ill-defined (Fig.1b). The boundary condition for the tangential magnetic field $\hat{\mathbf{z}} \times (\mathbf{H}_\parallel^{(1)} - \mathbf{H}_\parallel^{(2)}) = \sigma \mathbf{E}_\parallel$ exemplifies such a discontinuity. Here, $\mathbf{E}_\parallel$ and $\mathbf{H}_\parallel$ are the tangential (to the graphene surface) component of the electric and magnetic fields, respectively, and $\sigma$ is the surface conductivity of graphene. The difference between $\mathbf{H}_\parallel^{(1)}$ and $\mathbf{H}_\parallel^{(2)}$ is inconsequential only when the surface conductivity $\sigma$ is much smaller than the characteristic admittance[43–45] of the surrounding medium $Y_1$ (see Supporting Info S1 and S2). For graphene, this condition is satisfied in near-infrared and visible wavelengths (see Supporting Info S3). Here, the number in the subscripts and superscripts labels the media according to Fig. 1.

When graphene's surface conductivity is small, from the Coulomb-Lorentz force we show that the optical force, in a general 1D system, is simply given by the admittance of the wave in the absence of graphene (see the Methods section):

$$F_z(h) = \frac{F_{PEC} |\mathbf{E}_\parallel(h)/\mathbf{E}_{\parallel,\text{inc}}|^2}{2|Y_1|^2} \left[\text{Re}\{\sigma\}\text{Re}\{Y_{wave}(h)\} + \text{Im}\{\sigma\}\text{Im}\{Y_{wave}(h)\}\right], \qquad (1)$$

Here, $\mathbf{E}_{\parallel,\text{inc}}$ is the tangential component of the incident electric field, $\mathbf{E}_{\parallel}(h)$ is the tangential component of the electric field in the absence of graphene, calculated at the graphene position, and $F_{PEC}$ is the force exerted on a perfect mirror by the same plane wave (used for normalization). The wave admittance $Y_{wave}(h)$ is defined as the ratio of the tangential magnetic field over the tangential electric field[46], $\hat{\mathbf{z}} \times \mathbf{H}_{\parallel} = Y_{wave}(h) \mathbf{E}_{\parallel}$, before introducing graphene to the system. Note that the wave admittance includes the effect of the substrate, and is generally a space-variant complex value (see Supporting Info S2). Since the graphene sheet only has one degree of freedom $h$ in this 1D optical system, only the z-component of the optical forces is of interest.

In the particular case of a graphene suspended on a substrate with a complex-valued reflectivity $r = |r|\exp(i\phi_r)$ that represents the wave admittance given in Supplement S2, the optical force, to the first order in $\sigma/Y_1$, is given by

$$\frac{F_{Gr}}{F_{PEC}} = \frac{1}{2Y_1}\left[(1-|r|^2)\operatorname{Re}\{\sigma\} + 2|r|\operatorname{Im}\{\sigma\}\sin(2hk_z^{(1)} - \phi_r)\right] \equiv F_{absorption} + F_{scattering}(h), \qquad (2)$$

where $k_z^{(1)}$ is the z component of the wavenumber in vacuum.

Eq. (2) reveals that the optical forces on graphene have two distinct origins. The first term is related to the optical absorption of graphene, and is proportional to the real part of the surface conductivity $\operatorname{Re}\{\sigma\}$, i.e. the ohmic loss. This absorption force $F_{absorption}$ dominates when $r$ vanishes, as in the case of a freestanding graphene. On the other hand, the second term stems from optical scattering of graphene, and is associated with the imaginary part of the surface conductance $\operatorname{Im}\{\sigma\}$, i.e. the kinetic inductance. Note that the kinetic inductance ($\operatorname{Im}\{\sigma\}$), unlike the always-positive ohmic loss, can change sign depending on the frequency (See Supporting Info S3). This term dominates in systems with highly reflective substrates, for example, a prefect electrical conductor (PEC). In this latter case, $r \to 1$ and the wave admittance in Eq. (1) is purely imaginary. A main difference that sets the scattering force apart from the absorption force is that the scattering force $F_{scattering}(h)$ oscillates periodically between positive and negative values with respect to height $h$.

In the general case of a partially reflective substrate, both absorption and scattering forces contribute to the total force. The absorption (loss) leads to a constant downward pushing force, while the scattering force enables a tuning range, in which the total optical forces can be adjusted by changing the height of graphene from the substrate. The extent of the tuning is determined jointly by the reflectivity of the substrate and the kinetic inductance of graphene (the imaginary part of the graphene surface conductivity). And large values on both $|r|$ and $\text{Im}\{\sigma\}$ are necessary to support upwards pulling forces on graphene, i.e. the pulling scattering force to overcome the always-pushing absorption force at certain $h$ values. Eq. (2) also reveals the height of graphene at which the overall force reaches maximum or minimum, a quantity affected by the wavelength of the incident light and the phase of the reflectivity of the substrate $\phi_r$.

**Optical Forces under Strong Perturbation**

Below 20 THz, graphene alters the surrounding fields much more significantly with stronger absorption and scattering, resulting in larger optical forces. Although the perturbative treatment discussed above is no longer accurate, the insight we developed above still applies: optical forces on graphene is the sum of a height-independent term $F_{absorption}$ and a height-dependent term $F_{scattering}(h)$, which are related to the traveling waves and the standing waves in the system, respectively. Therefore, in this section we investigate the overall evolution of optical forces on graphene through a wider spectral range from 0.1 to 300 THz.

In general, both traveling waves and standing waves exist in this optomechanical system (Fig. 1), as the graphene sheet and the substrate serves as the two partial mirrors of an asymmetric Fabry-Perot microcavity. A traveling wave is associated with the total transmission, and has a constant strength throughout. The standing waves, in contrast, only exist above the substrate and are of different strengths above and below graphene. Because the frequency and the quality factor of the asymmetric Fabry-Perot resonance depend on three parameters: the suspension height, the reflectivity of the graphene (as a function of the optical frequency), and the reflectivity of the substrate (as a function of its permittivity), we next systematically study their effects.

**1. Fixed suspension height and varied substrate permittivity**

We choose a fixed separation $h$ of 200 nm, and vary the substrate permittivity from 1 to 100 to illustrate the transition from predominantly traveling-wave environments to predominantly standing-wave environments. This level of separation has been used in experiments[42,47] that measure the mechanical resonance of graphene membranes. For a purely standing-wave environment (substrate made of $\epsilon_3 = -\infty$, perfect electrical conductor), the exact solutions (solid curves) from the Maxwell stress tensor agree well with the first-order perturbation results (dashed curves) obtained via Eq.(2), as shown in Figs.2b and 2c. Above 20 THz, as predicted by the first-order perturbation theory [Eq.(2)], optical forces follow the variation of ohmic loss for low-index substrates (poorly reflective), while the kinetic inductance dictates the optical forces when high-index substrates (highly reflective) are used. For example, the optical forces spectrum (red) without a substrate ($\epsilon_3 = 1$) closely resembles the real part of the graphene conductivity (Fig. S2 of the Supporting Info), with a step increase at $2E_F$ and a monotonic increase with decreasing frequencies. As the permittivity of the substrate rises, reflection and the associated standing-wave components become more significant in the total optical forces. Thus the sign change in the imaginary part of graphene conductivity drives a similar sign change in the optical forces, from a pushing force to a pulling force at 182 THz. This transition is most pronounced in the case of a PEC substrate, where the optical forces (the black curve in Figs. 2b and 2c) strongly resemble the imaginary part of the conductivity (Fig. S2).

In comparison to the high-frequency perturbative regime (Fig.2c), the low-frequency regime (Fig.2b) yields much larger forces. The optical forces plateau below 1 THz, following the plateaus of the reflection and absorption (Fig S2c). Under such fixed separation, the maximum force occurs without a substrate, *i.e.*, $\epsilon_3 = 1$. The amplitude of the force (~33% of $F_{PEC}$) agrees well with the ~46% power absorption and ~10% reflection (Fig.S2), suggesting that the optomechanical response of graphene to plane-wave incidence at microwave frequencies is dominated by the absorption of the momentum of the incident photons, with a smaller contribution from reflection. In contrast, for highly-reflective substrates, as frequency falls, the fixed 200nm separation is much smaller than the free-space wavelength, and the graphene sheet is located closer to a nodal plane of the E fields. The amplitude of the tangential electric fields near graphene vanishes, and the induced current on graphene and the resultant optical forces also vanish. To get a large optical force, one needs to lift graphene away from the substrate to a height of $h \sim \pi / 4k_z^{(1)}$.

Higher reflectivity from the substrates generally translates to larger optical force (Fig. 2d) in the perturbative regime above 20THz. This enhancement of more than an order of magnitude is due to a low-*Q* Gires–Tournois etalon[48–50] (asymmetric Fabry-Perot resonator) formed between the highly reflective substrate and the weakly reflective graphene 'mirror'. This enhancement is also robust over a range of doping levels ($E_F \approx 0.2 - 0.45$eV) as shown in Fig. 2e. However, the absolute magnitude of the optical forces in this frequency range is small, far below the maximum possible values we will discuss in the next section.

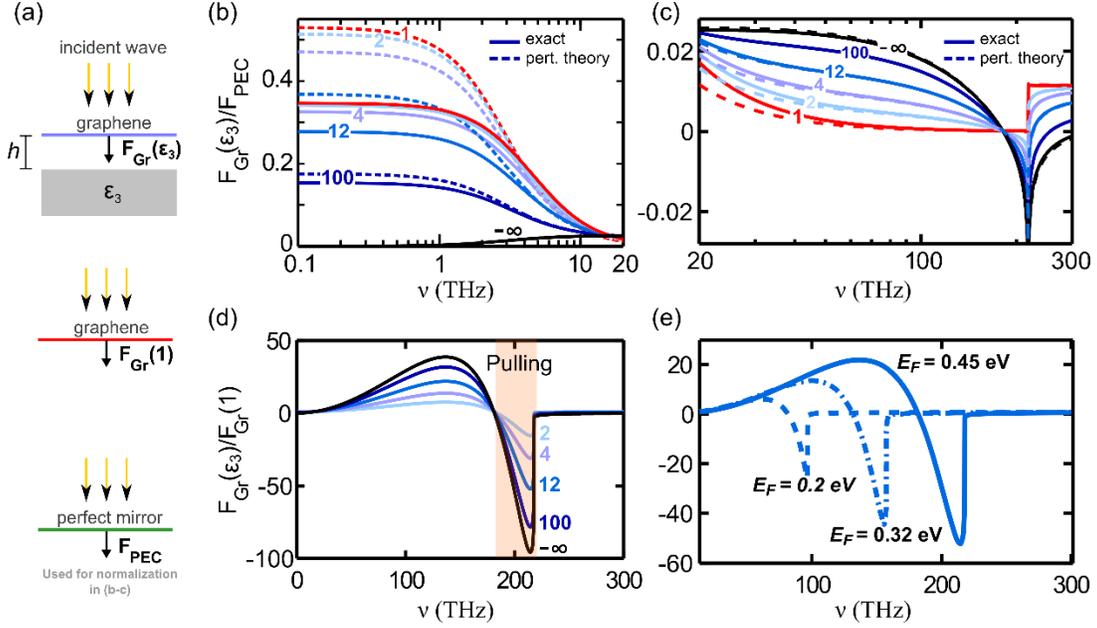

Figure 2. Normal-incidence optical forces on graphene suspended 200 nm above various substrates. (a) Schematics of the systems: (top) graphene above a substrate with permittivity $\epsilon_3$ (middle) free-standing graphene, and (bottom) a perfect mirror as a force reference. (b) Optical force on graphene in THz and far-IR regime, and (c) optical forces in the perturbative mid-IR regime. The exact results (solid curves) agree with the perturbative first-order approximation (dashed curves) qualitatively at low frequencies and quantitatively at high frequencies. (d) Enhancement of optical forces: graphene on a substrate vs. free-standing graphene. (e) The effect of the Fermi levels on the optical forces with a substrate permittivity of 12. In all panels except for panel e, the Fermi level is at 0.45 eV, and curve colors represent substrate permittivity as indicated in the legend of the panel c. In all the figures, the graphene scattering rate $\Gamma$ is chosen to be 3 THz.

With such a sub-wavelength separation, pulling (repulsive) forces that elevate graphene away from the substrate can be found in a narrow frequency range around 200 THz, right below the onset of the interband transition (Fig.2d). In this frequency range, the graphene kinetic inductance is positive, and graphene behaves as an anisotropic dielectric surface (see Fig. S2). Being proportional to the kinetic inductance, the optical force has the opposite sign to that in the low-frequency regime, and pulls the graphene *towards* the incident fields. This pulling force agrees with the fact that high-index dielectric generally moves towards the maxima of the electric field, one of which is located approximately quarter wavelength above a high-index substrate. Therefore, the elevating force occurs when graphene is closely placed to such substrate surface, satisfying $h < \pi / 2k_z^{(1)}$ or $h < \lambda / 4$ at normal incidence.

## 2. Varied suspension height with a silicon substrate

With a highly reflective substrate, the scattering force becomes dominant. Using a silicon substrate ($\epsilon_3 = 12$) as an example, we calculate optical forces on a graphene layer (E$_F$=0.45eV) with the separation $h$ adjusted between 0 and 1.5 $\mu m$ and the frequency varied between 0.1 and 300 THz (Fig. 3). All the calculations are performed using the Maxwell stress tensor. The graphene layer and the underlying substrate act as two partial mirrors, forming a low-finesse asymmetric Fabry-Perot cavity, and the total force becomes periodic with respect to the separation $h$. The constant-force contours (Fig. 3a) follow the general trend of $\nu \propto 1/h$, with minor modification from the dispersive surface conductivity (See Supporting Info S3). The zeros of the force (dashed lines in Fig. 3b) occurs at $h$ values that satisfy the relation $2hk_0 = m\pi$. The optical force at those separations vanishes because the tangential electric field $\propto \sin(hk_z^{(1)})$ vanishes for even values of $m$, or the tangential magnetic field $\propto \cos(hk_z^{(1)})$ vanishes for odd values of $m$. In general, illuminating this asymmetric Fabry-Perot cavity at an off-resonance frequency $\omega$ induces an optical force on the graphene.

The separation leading to zero optical forces is either a stable or unstable equilibrium, depending on the sign of the force gradient. The highly dispersive nature of graphene conductivity creates three separate frequency regimes (Fig. 3b). Above the onset of the interband transitions ($h\nu = 2E_F$), the optical absorption dominates, and the total force is a small downward pushing force regardless of height with fluctuating amplitude, with no equilibrium height. In the spectral range (shaded regions in Fig. S2) between the onset of the interband transitions and the frequency of zero kinetic inductance Im$\{\sigma\}$ (the horizontal dotted line in Fig. 3b), graphene acts as a dielectric surface, and the stable heights occur at odd values of *m*. Further lowering the frequency, graphene acts as a plasmonic surface (Im$\{\sigma\} < 0$), and the stable equilibrium positions shift to even values of *m*. Although this regime provides the largest optical forces and the most stable equilibrium among all three regimes, the associated stable height is at least 700 nm above the substrate, suggesting a challenge of large DC voltage required in electrostatic doping. The locations of a stable equilibrium is consistent with the fact that in a standing-wave system, a plasmonic sheet is stable at the zeros of the electric field and the maxima of the magnetic field, while a dielectric film prefers to be located at the maxima of the electric field and the zeros of the magnetic field. An incoming power 6.8 $\mu W$ at the frequency $v$=3 THz (which corresponds to the maximal forces) creates a 1eV-deep potential well (assuming the graphene is doped to $E_F$=0.45 eV).

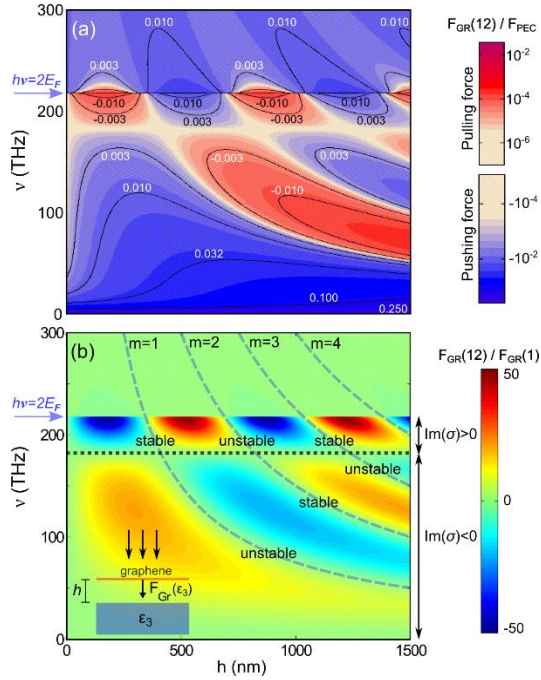

Figure 3. Optical forces on graphene ($E_F$=0.45eV) at normal incidence as a function of frequency $v$ and the separation $h$ from a silicon substrate with $\epsilon_3 = 12$. (a) Optical forces normalized to $F_{PEC}$ (twice the incoming photon momentum flux). (b) Optical forces normalized to that on a free-standing graphene.

One can find the maximum of the optical spring constant, at the locations of such stable equilibrium. The optical spring constant is an important metric in optomechanics to compare the stability of an optical trap. In the high-frequencies perturbative regime, it is given by: $\nabla_h F_z^{[1]} = 2k_z^{(1)} F_{PEC} |r| \operatorname{Im}\{\sigma\} \cos(2hk_z^{(1)} - \phi_r) / Y_1$. This quantity is crucial in a range of applications, such as ground-state cooling.[42,51,52]. The optical spring constant reaches its maximum of $2k_z^{(1)} F_{PEC} |\mathrm{r}| \mathrm{Im}\{\sigma\}/Y_1$ at the center of the optical potential well. In comparison to other optomechanical systems[53], the optical spring constant of graphene appears small. However, considering the low mass density, the resulting acceleration can be significant, which may be exploited in graphene structures with low in-plane stress.

## 3. Tuning range of optical forces for various substrates

Since optical forces on graphene are periodic with respect to height *h*, we summarized the range of optical forces (Figs.4a-b, e-f) from 0.1-300 THz for a range of substrates: low-index dielectric substrates ($\epsilon_3 = 1$ or 2.25), high-index (silicon) substrates, and several common plasmonic substrates ($\epsilon_3 < 0$). The maximum possible pulling forces (dashed curves) and pushing forces (solid curves) define a tunable range (the shaded region in Fig. 4). The extent of the tuning is determined jointly by the reflectivity of the substrate and the kinetic inductance of graphene. The largest range occurs when the optical frequency is identical to the carrier scattering rate of graphene and when one uses highly reflective (high-index or metallic) substrates. We found that the presence of losses and dispersion in metals (Fig. 4d) barely modifies the tuning range of the forces (Fig. 4a). However, this range diminishes by using lower index substrates with small reflectivity, or by operating at the high-frequency limit. The mid-point of this range is dictated by the optical absorption of graphene, and therefore follows the spectral features of the real part of the graphene surface conductivity (Fig. S2). Weakly reflective substrates result in a very small tuning range with the optical forces largely following an absorption-dependent mean value (the blue curve in Fig. 4a), which peaks at DC frequency. Therefore, for highly reflective substrates, the absolute maximum of the pushing force is found when optical frequency equals the free-carrier scattering rate of graphene, and for low-index substrates, the maximum occurs at DC frequency. Generally, the presence of a reflective substrate enhances the amplitude of the optical force with respect to that of a freestanding graphene (Figs. 4b, 4f).

Both the operational frequency and the substrate permittivity also influence strongly the separation heights at which the force maxima and minima are located. In the perturbative regime, peak forces are found where the electric and magnetic fields are equal in magnitude and 90 degrees out of phase. For dielectric substrates and plasmonic substrates away from their plasmon frequencies, this condition is satisfied in Eq. (2) when *h* is a solution of $\sin(2hk_0) = \pm 1$. For graphene in the plasmonic regime ($\text{Im}\{\sigma\} < 0$), the maximum pushing force is found at the separation $h_{max} = \lambda/8$ (i.e., $\sin(2hk_0) = +1$) and the maximum pulling force is found at $h_{min} = 3\lambda/8$ (i.e., $\sin(2hk_0) = -1$). Since the absorption force dominates at low frequencies and is always pointing downwards, we instead use the term "minimum force" $F_{min}$ for lower bound of the range for generality. For graphene in the dielectric regime ($\text{Im}\{\sigma\} > 0$), the exact opposite occurs: $h_{max} = 3\lambda/8$ and $h_{min} = \lambda/8$ (Fig. 4c). This reversal near 180 THz is accompanied by vanishing $F_{max}$ and $F_{min}$ observed in Figs. 4b and 5e. In the strong perturbation limit of 20THz and below, we also need to take into account the finite phase shift from the reflection of a lossy substrate. Thus, the separation associated with the maximum/minimum forces are frequency dependent (see Fig. 4c). The change in separation due to the reflectivity phase different from $\pi$ can be seen from Eq. (2) as $h_{max} = (3/8 + \phi_r/4\pi)\lambda$ in the plasmonic regime and $h_{max} = (5/8 + \phi_r/4\pi)\lambda$ in the dielectric regime. This change is particularly pronounced at high frequencies when the real part and the imaginary part of the substrate permittivity is comparable at near-infrared frequencies for metal (compare the orange and black curves in Fig. 4c). In the non-perturbative regime ($\nu < 10\,\text{THz}$), graphene surface conductivity becomes predominantly real-valued (see Fig. S2), shifting the maximum force gradually to $h_{max} = \lambda/4$, and the minimum force gradually to $h_{min} = \lambda/2$.

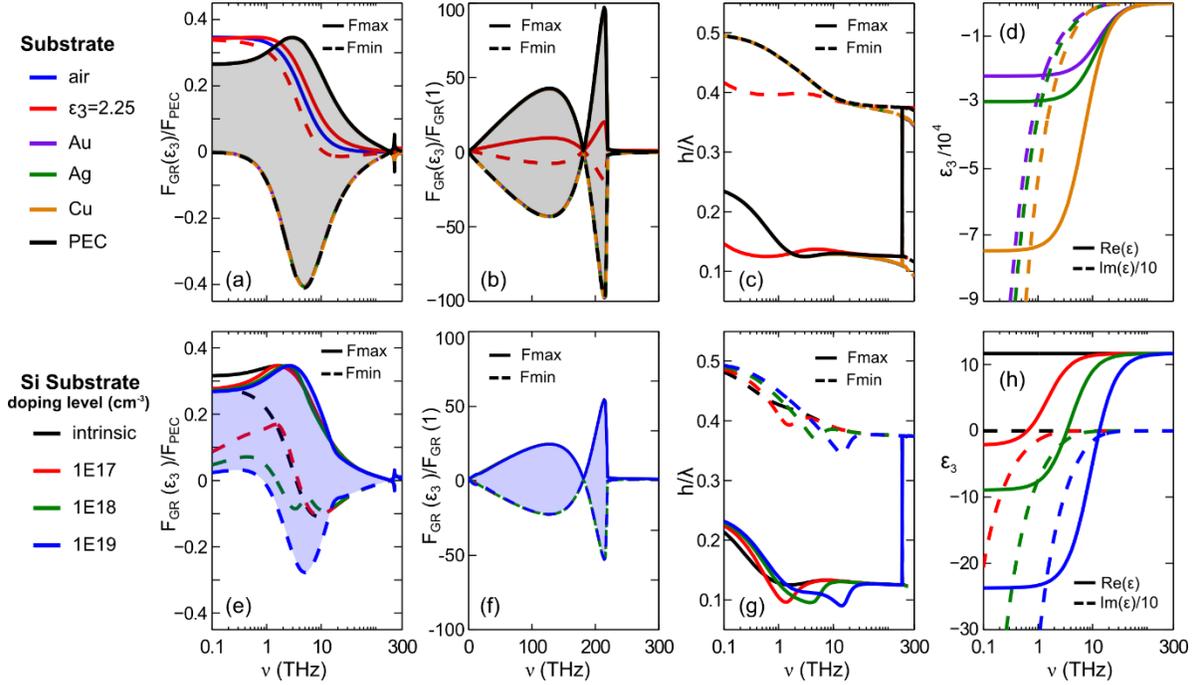

Figure 4. (a-b, e-f) Maximum (solid lines) and minimum (dashed lines) optical forces on suspended graphene ($E_F$=0.45eV) for various substrate materials (shown in the legend to the left of the panels), normalized by the forces on a PEC mirror (panel a and e) and by the forces on a free-standing graphene (panel b and f) (c,g) The graphene-substrate separation where $F_{max}$ and $F_{min}$ are found. (d,h) The permittivity of the substrates. The plasma frequencies and carrier scattering rates in silicon [in unit of THz] for various doping levels $n$ are: $(n = 10^{17} \text{cm}^{-3}, \omega_p = 5.57, \Gamma = 1.50)$, $(n = 10^{18} \text{cm}^{-3}, \omega_p = 17.62, \Gamma = 3.88)$, and $(n = 10^{19} \text{cm}^{-3}, \omega_p = 55.72, \Gamma = 9.37)$.

Silicon substrate, commonly used for electrostatic gating, requires particular care in treating the optical forces, because its reflectivity can vary strongly. Silicon possesses a plasmon frequency inside the mid- and far-infrared frequencies, depending on its doping level. The corresponding optical forces are plotted in Figs. 4 (e-f). The doped silicon is modeled using a Drude-like permittivity[54,55]: $\epsilon_{Si}(\omega)/\epsilon_0 = 3.415^2 - \omega_p^2 /[\omega(\omega - i\Gamma)]$, where $\omega_p = \sqrt{ne^2/(m^*\epsilon_0)}$ is the plasma frequency. $\Gamma = e/(m^*\mu)$ is the scattering rate, and $\mu$ is the electron mobility, a function of the doping level $n$ [56]. The effective electron mass $m^*$ is assumed to be 0.26 that of the electron mass $m_0$. As a reference, we also plot the case for intrinsic silicon (black curve). Figs. 4(e-f) suggest that higher doping levels lead to higher reflectivity, thereby increasing the tunable range of the optical forces on graphene. When doping level reaches $10^{19}$ cm$^{-3}$, the optical force spectrum resembles the one obtained from a metallic substrate (see Fig. 4a). The separation associated with the maximum and minimum forces also depend on the doping level (Fig.4g). It is noteworthy that the separation for the maximum force reduces to a level (~0.08$\lambda$) much smaller than that of other highly reflective substrates near the transparency threshold of silicon ($\epsilon_{Si}(\omega) \sim 0$), since there the phase of the reflectivity abruptly changes.

Extending from the normal incidence illumination to the more general oblique incidence, together with the incorporation of the finite temperature effects, the analytical results derived above still apply. For oblique incidences, one needs to consider three major differences: The equilibrium spacing, shown in Fig. 3, becomes larger and inversely proportional to the cosine of the incident angle, since $k_z$ is reduced in Eq. (2) for an oblique incidence. The absolute value of the normal force $F_z$ decreases with an increasing incidence angle, similar to the decreased force experienced by a perfect mirror: $F_{PEC} = 2P/c \cdot \cos^2 \theta$, where $P$ is the incoming power (irradiance), and $\theta$ is the incidence angle. Additionally, an in-plane optical force is exerted on the graphene, caused by the in-plane momentum of the photons lost in optical absorption. On the other hand, taking into account the effect of finite temperature, the logarithmic divergence of the surface conductivity is smeared out at the onset of the interband transition (Supporting Info S3). At this frequency range, the optical forces are already comparatively small, and such smearing will further reduce the forces. However, at lower frequencies from DC to tens of THz, at which the optical forces on graphene are large enough to be useful, such smearing has little effect on the surface conductivity and the reported optical forces (see Fig. S3 and S4). Also note that by increasing the temperature, only the cases in which optical forces depend mainly on the absorption (i.e. the case of low-index substrates) are affected the most.

In summary, we investigated conditions to create large optical forces on single-layer graphene suspended above a reflective substrate at THz and infrared wavelengths. Large optical forces require the frequency of the illumination to be near the free-carrier-scattering rate of graphene, and the peak force is comparable to 40% of the scattering force experienced by a perfect mirror. The overall optical force consists of two main components: an absorption force largely determined by the real part of the surface conductivity of graphene, and a scattering force largely determined by the kinetic inductance (the imaginary part of the surface conductivity). The absorption force is independent of the separation height, whereas the scattering force changes both its direction and amplitude as a periodic function of the separation, thanks to a low-$Q$ Fabry-Perot etalon formed between the graphene 'mirror' and the substrate. The absorption force dominates with a weakly reflective substrate, while the tunable scattering force dominates when a highly reflective substrate is used. For graphene, the large tuning range of the scattering force and the possibility of optical levitation enable a variety of optomechanical applications.

**Methods**

In high frequencies, graphene behaves as a small perturbation and the discontinuities in the fields $\hat{\mathbf{z}} \times (\mathbf{H}_\parallel^{(1)} - \mathbf{H}_\parallel^{(2)}) = \sigma \mathbf{E}_\parallel$ diminish. The fraction of the surface conductivity to the characteristic wave admittance of the medium measures such smallness. When this ratio is small, the fields can be expanded in terms of the perturbation ratio $\sigma/Y_1$:

$$\mathbf{E} = \mathbf{E}^{[0]} + \frac{\sigma}{Y_1}\mathbf{E}^{[1]} + \dots, \quad \mathbf{H} = \mathbf{H}^{[0]} + \frac{\sigma}{Y_1}\mathbf{H}^{[1]} + \dots \ .$$

$\mathbf{E}^{[0]}$ and $\mathbf{H}^{[0]}$ are the fields in the absence of graphene ($\sigma = 0$), and numbers in brackets '[]' refer to the order of the expansion. To the first order, the optical force can be integrated from the Coulomb-Lorentz force density over the entire graphene surface $S$:

$$F_z^{[1]} = \frac{1}{2}\mathrm{Re}\left\{\int_S \left[\rho^{[1]} E_z^{[0]*} + \mu_0(\mathbf{J}^{[1]} \times \mathbf{H}_\parallel^{[0]*})\cdot\hat{\mathbf{z}}\right]da\right\}.$$

$\rho$ denotes the surface charge density, and $\mathbf{J}$ denotes surface current density in graphene. Both vanishes at the 0th order, e.g., $\mathbf{J}^{[0]} = 0$. Using the force exerted on a perfect mirror by the same plane wave $F_{PEC}$, we normalize such an optical force to the incident power as

$$F_z^{[1]} = -F_{PEC}\frac{\left(\mathbf{J}^{[1]} \times \mathbf{H}_\parallel^{[0]*} + \mathbf{J}^{[1]*} \times \mathbf{H}_\parallel^{[0]}\right)\cdot\hat{\mathbf{z}}}{4Y_1^2\,|\mathbf{E}_{\parallel,\mathrm{inc}}|^2}\ .$$

Here, $\mathbf{E}_{\parallel,\mathrm{inc}}$ is the amplitude of the tangential component of the incident electric field. Note that the unperturbed fields are continuous, inducing a surface current $\mathbf{J}^{[1]} = \sigma \mathbf{E}_\parallel^{[0]}$.

To track down the main contributing factors to the optical force, we further express the tangential magnetic field using the tangential electric field: $\hat{\mathbf{z}} \times \mathbf{H}_\parallel^{[0]} = Y_{wave}(h)\,\mathbf{E}_\parallel^{[0]}$. Here, the wave admittance $Y_{wave}(h)$ is defined as the ratio of the tangential magnetic field over the tangential electric field[46], before introducing graphene to the system. Note that the wave admittance includes the effect of the substrate, and is generally a space-variant complex value (see Supporting Info S2). Thus, the expression for the optical force becomes $h$-dependent:

$$\begin{aligned}F_z^{[1]}(h) &= \frac{F_{PEC}\,|\mathbf{E}_\parallel^{[0]}/a_1|^2}{2|Y_1|^2}\mathrm{Re}\left\{\sigma Y_{wave}^*(h)\right\} \\ &= \frac{F_{PEC}\,|\mathbf{E}_\parallel^{[0]}/a_1|^2}{2|Y_1|^2}\left[\mathrm{Re}\{\sigma\}\mathrm{Re}\{Y_{wave}(h)\} + \mathrm{Im}\{\sigma\}\mathrm{Im}\{Y_{wave}(h)\}\right]\end{aligned}, \quad (3)$$

Note that in the phasor convention used in this paper [$\exp(i\omega t)$], the imaginary part of both the surface conductivity and the wave admittance is opposite to those in the other common convention [$\exp(-i\omega t)$]. Therefore, the force expression given in Eq. (3) is independent of the convention used.

## ASSOCIATED CONTENT

**Supporting Information**

The Supporting Information includes (1) Exact Optical Force Experienced by a Graphene Layer Suspended on a Substrate, (2) Wave Admittance of the Space above a Substrate, (3) Optical Properties of Graphene, (4) Temperature Effect on Optical Forces, and (5) Comparison between Optical Forces and Electrostatic Forces. This material is available free of charge via the Internet at http://pubs.acs.org.

## AUTHOR INFORMATION

Corresponding Author
*E-mail: hmousavi@gmail.com
†E-mail: zheng.wang@austin.utexas.edu

Notes

The authors declare no competing financial interest.

## ACKNOWLEDGEMENTS

We would like to thank Alexander Khanikaev and Gennady Shvets for many helpful comments. This work is in part supported by the Packard Fellowships for Science and Engineering, and the Alfred P. Sloan Research Fellowship.